\documentclass{iopart}
\usepackage{graphicx}
\usepackage{dcolumn}
\usepackage{bm}
\usepackage{color}
\newcommand{\lucu}{LuCuGaO$_4$ }
\newcommand{\lucuns}{LuCuGaO$_4$}
\usepackage{amssymb}
\usepackage{epstopdf}
\begin{document}

\title[Investigations of the low temperature state of \lucuns]{Neutron scattering and $\mu$SR investigations of the low temperature state of \lucuns}
\author{S. Calder$^1$\footnote{Began this work at London Center for Nanotechnology, 17-19 Gordon Street, London WC1H 0AH, UK}, SR Giblin$^2$\footnote{Began this work at ISIS Facility, Rutherford Appleton Laboratory, Chilton, Didcot, Oxfordshire OX11 0QX, UK}, DR Parker$^3$\footnote{Now at  Oxford Instruments, Tubney Woods, Abingdon, Oxfordshire, OX13 5QX, UK}, PP Deen$^4$\footnote{Also at Niels Bohr Institute, Universitetsparken 5, 2100 Copenhagen, Denmark; began this work at Institut Laue Langevin,  BP 156, 6, rue Jules Horowitz, 38042, Grenoble Cedex 9, France}, C Ritter$^5$, JR Stewart$^6$\footnote{Began this work at Institut Laue Langevin,  BP 156, 6, rue Jules Horowitz, 38042, Grenoble Cedex 9, France}, S Rols$^5$, T Fennell$^7$\footnote{Began this work at London Center for Nanotechnology, 17-19 Gordon Street, London WC1H 0AH, UK; and  Institut Laue Langevin,  BP 156, 6, rue Jules Horowitz, 38042, Grenoble Cedex 9, France.}}

\address{$^1$Quantum Condensed Matter Division, Oak Ridge National Laboratory, Oak Ridge, TN 37831, USA.}
\address{$^2$School of Physics and Astronomy, University of Cardiff, Cardiff, CF24 3AA, UK}
\address{$^3$Department of Chemistry, University of Oxford, Inorganic Chemistry Laboratory, South Parks Road, Oxford, OX1 3QR, UK}
\address{$^4$European Spallation Source ESS AB - Box 176, 22100 Lund, Sweden}
\address{$^5$Institut Laue Langevin,  BP 156, 6, rue Jules Horowitz, 38042, Grenoble Cedex 9, France}
\address{$^6$ISIS Facility, Rutherford Appleton Laboratory, Chilton, Didcot, Oxfordshire OX11 0QX, UK}
\address{$^7$Paul Scherrer Institut, 5232 Villigen PSI, Switzerland}
\ead{tom.fennell@psi.ch}

\begin{abstract}
LuCuGaO$_4$ has magnetic Cu$^{2+}$ and diamagnetic Ga$^{3+}$ ions distributed on a triangular bilayer and is suggested to undergo a spin-glass transition at $T_g\sim 0.4$ K.  Using $\mu$SR and neutron scattering measurements, we show that at low temperature the spins form a short-range correlated state with spin fluctuations detectable over a wide range of timescales: at 0.05 K magnetic fluctuations can be detected in both the $\mu$SR time window and also extending beyond 7 meV in the inelastic neutron scattering response, indicating magnetic fluctuations spanning timescales between $\sim10^{-5}$ and $\sim10^{-10}$ seconds.  The dynamical susceptibility scales according to the form $\chi''(\omega)T^\alpha$, with $\alpha=1$, throughout the measured temperature range ($0.05 - 50$ K).  These effects are associated with quantum fluctuations and some degree of structural disorder in ostensibly quite different materials, including certain heavy fermion alloys, kagome spin liquids, quantum spin glasses, and valence bond glasses.  We therefore suggest \lucu is an interesting model compound for the further examination of disorder and quantum magnetism.

\end{abstract}

\maketitle

\section{\label{sec:Introduction}Introduction}

\begin{figure}
	\centering
	\includegraphics[trim=40 520 110 80,clip=true,scale=0.5]{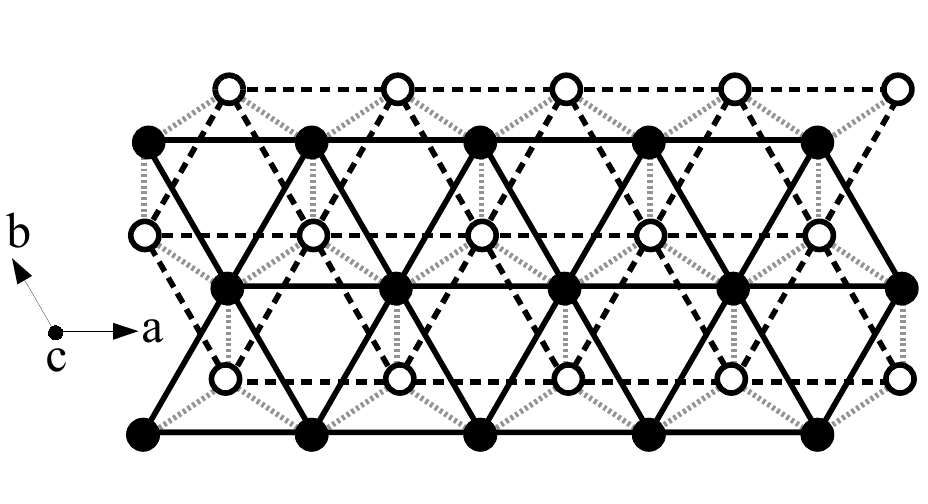}\\
	\includegraphics[trim=40 480 110 120,clip=true,scale=0.5]{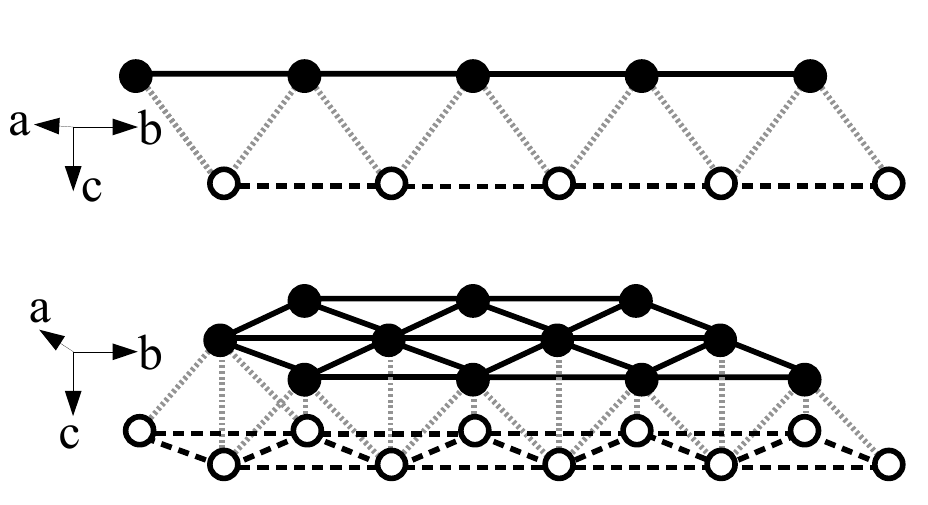}\\
	\caption{The geometry of the triangular bilayers formed by Cu$^{2+}$ and Ga$^{3+}$ cations in \lucuns.  The two faces of the bilayer are shown in black and white.   The shortest bonds (called $J_1$ in the text) are those that link the two faces of the bilayer forming a puckered honeycomb lattice (grey).  Although the honeycomb lattice is bipartite and cation order is therefore possible, the side views (middle and bottom) show that the puckering of this lattice means that such order would cause charge separation across the layer.  The bond distance within the faces of the bilayer are slightly longer and form triangular lattices (black and dashes) on which cation (and magnetic) order will be highly frustrated.}
	\label{fig:LuCuGaO4topology}
\end{figure}

\lucu~\cite{cava:1998} encompasses several topics of interest in condensed matter physics.  It has the same crystal structure as the multiferroic LuFe$_2$O$_4$~\cite{ikeda:2005}: both contain triangular bilayers (Figure~\ref{fig:LuCuGaO4topology}), a two-dimensional geometrically frustrated lattice which has so far not been widely studied.  Theories of resonating valence bond (RVB) states and other spin liquid ground states~\cite{normand:2009,Balents:2010p3111} typically address low-dimensional frustrated magnets with $S=1/2$ and experimental realizations are sought-after for comparison.  In \lucu the magnetic Cu$^{2+}$ ions have $S=1/2$,  suggesting the possibility of quantum magnetic phenomena, but are strongly diluted by diamagnetic Ga$^{3+}$ resulting in a material with quenched disorder and charge frustration~\cite{cava:1998,anderson:1956}.

At a classical level, spins with antiferromagnetic couplings residing on lattices with triangular geometry are frustrated: all pairwise interactions cannot be minimized simultaneously.  Other systems also map onto such models, for example the configurations of two cations on such a lattice are identical to those of antiferromagnetically coupled Ising spins on the same lattice~\cite{anderson:1956}, a type of charge frustration.  In LuCuGaO$_4$ the Cu$^{2+}$ and Ga$^{3+}$ cations are distributed over a single crystallographic site, forming well separated triangular bilayers.   A single bilayer contains two opposed triangular lattices (see Figure~\ref{fig:LuCuGaO4topology}), but the shortest distance between cations is between opposite faces of the bilayer, forming a puckered honeycomb lattice. The honeycomb and triangular geometries may be conveniently viewed as a $J_1-J_2$ model on the honeycomb lattice~\cite{fouet:2001,mattsson:1994}, which is also a frustrating geometry for magnetic order (see Figure~\ref{fig:LuCuGaO4topology}). \lucu is therefore expected to exhibit both charge and magnetic frustration.

The combination of a frustrating lattice geometry and $S=1/2$ is expected to lead to unconventional groundstates such as a spin liquid, as the energy can be further reduced below the classical groundstate by quantum fluctuations (types and definitions of spin liquids are reviewed by Normand in ~\cite{normand:2009} and Balents in ~\cite{Balents:2010p3111}).  Whilst geometrically frustrated magnets have typically been viewed as clean systems in which the effect of frustration can be isolated (in contrast to spin glasses, where structural disorder is also inevitable),  it has become apparent that weak structural disorder or non-magnetic impurities can determine the low temperature behaviour of  frustrated magnets, as any perturbation of a fragile groundstate is magnified in importance.  Theoretical examples include the pinning of dimers around non-magnetic impurities in the $S=1/2$ kagome Heisenberg antiferromagnet~\cite{rousochatzakis:2009,lauchli:2007,dommange:2003} or a strain-induced spin glass transition in the pyrochlore Heisenberg antiferromagnet with weak bond disorder~\cite{saunders:2007}, while experimental examples include the apparent sensitivity of the ground state of the rare earth titanates Yb$_2$Ti$_2$O$_7$~\cite{Yaouanc:2011bj,Ross:2012dj} and Tb$_2$Ti$_2$O$_7$~\cite{taniguchi} to tiny modifications of the stoichiometry.  In SCGO (SrCr$_{8-x}$Ga$_{4-x}$O$_{19}$), experimental evidence of spin clusters nucleated at diamagnetic defects has recently been interpreted as the formation of a fractional spin texture~\cite{mendelsscgoprl,limot,Sen:2011jt}.

\begin{figure}
	\centering
	\includegraphics[trim=300 290 450 230,clip=true,scale=0.2]{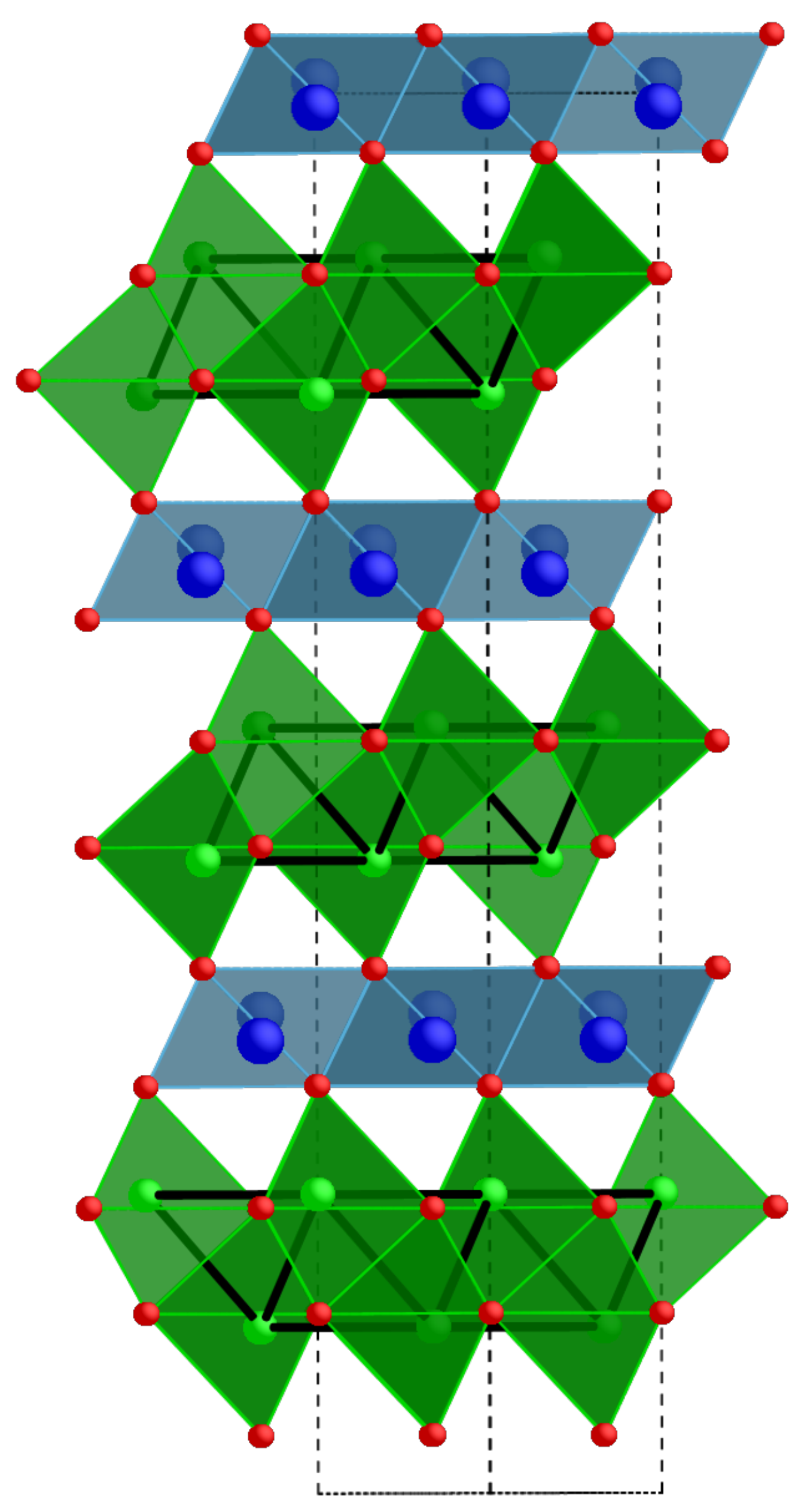}
	\caption{Crystal structure of LuCuGaO$_4$ as obtained from the Rietveld refinement shown in Figure~\ref{fig:LuCuGaO4_structure}.  The $c$-axis is vertical on the page, the projection axis is the $a-b$ diagonal.  The bilayers (black bonds) are formed of edge-sharing $M^{(')}$O$_5$ trigonal bipyramids ($M^{(')}$ are shown in green, O$^{2-}$ in red).  The bilayers are separated by layers of edge-sharing LuO$_6$ octahedra.  The Lu$^{3+}$ cations randomly occupy one of two sites displaced from the centre of each octahedron (blue).}
\label{fig:LuCuGaO4xtalstructure}
\end{figure}

\lucu is a member of the series $RMM'$O$_4$, where $R$ is a small rare earth such as Yb$^{3+}$ or Lu$^{3+}$, $M$ is a late $3d$ transition metal cation with $+2$ charge and $M'$ is a $+3$ cation from the late transition metals or main group.  Much attention has been focussed on RFe$_2$O$_4$, where charge ordering of Fe$^{2+}$ and Fe$^{3+}$ is possible, and is thought to underlie the multiferroicity in LuFe$_2$O$_4$~\cite{ikeda:2005,Mulders:2011fs,deGroot:2012fi}, while in YbFe$_2$O$_4$ it appears that the charge frustration drives the formation of an incommensurate charge density wave~\cite{Hearmon:2012ih}.  However, numerous examples with two different cations have also been characterized.   As well as \lucuns; YMnFeO$_4$, Yb$M$FeO$_4$ ($M$ = Mg, Fe, Co, Cu), YbCuGaO$_4$, Lu$M$FeO$_4$ ($M$ = Zn, Fe, Co, Cu) and LuCoGaO$_4$ have all been studied~\cite{cava:1998,wiedenmann:1983,iida:1990,iida:1990:2,isobe:1990,tanaka:1990,iida:1991,matsumoto:1992,ikeda:1995,tanaka:1995,todate:1997,todate:1998:jpcm,todate:1998:prb,sunaga:2001,dabkowska:2002,yoshii:2007,fritsch:2012}.  Structurally, all are described as having randomly distributed $M$ and $M'$ ions on the bilayer (the crystal structure is illustrated in Figure~\ref{fig:LuCuGaO4xtalstructure} and described in greater detail below).  With the exception of $R$CuGaO$_4$ ($R$ = Yb, Lu)~\cite{cava:1998} all these materials exhibit a splitting of field cooled and zero field cooled magnetic susceptibilities at temperatures of order 20 K (and in some cases considerably higher), indicative of freezing or spin glass transitions.  Previously the most extensively investigated stoichiometry was LuMgFeO$_4$~\cite{wiedenmann:1983,ikeda:1995,tanaka:1995,todate:1997,todate:1998:prb}.  Studies using magnetization measurements, neutron scattering and M\"ossbauer spectroscopy established that there are short range magnetic correlations associated with frustrated two-dimensional ordering and suggested that random dilution leads to the formation of clusters and two types of site: those lying in the body of clusters and having a full complement of magnetic neighbors, and those lying on extended branches of clusters and having just one or two neighbors.

Studies of \lucu also suggest that the copper and gallium ions are randomly distributed on the sites of the bilayer and consequently that  \lucu is both spin and charge frustrated~\cite{cava:1998}.  Indeed, despite a Curie-Weiss temperature of $\theta_{CW} = -69$ K, no magnetic ordering or freezing transition is observed down to 0.4 K where a broad peak in the ac-susceptibility is currently attributed to a spin glass transition \cite{cava:1998}.   

The primary difference between those $RMM'$O$_4$ with high temperature freezing transitions and \lucu is the presence of (only)  $S=1/2$ moments.  In Table~\ref{weisstempstab} we list frustration indices for compounds with diamagnetic rare earth and a single magnetic ion on the triangular bilayer.  If the conventional frustration index ($\theta_{CW}/T_g$) is normalized to the size of the moment, we see that \lucu is considerably more frustrated than the large moment systems.  Since the systems typically are interpreted as freezing (hence we employ $T_g$ not $T_N$ in our frustration indices), we suggest that fluctuations are considerably more important in \lucuns.  In this paper we use polarized neutron scattering, $\mu$SR and neutron spectroscopy to produce a detailed microscopic characterization of LuCuGaO$_4$ which shows that a correlated state with fluctuations on many timescales exists, even in the presence of strong structural disorder.  This state contrasts with a classical spin glass, where the fluctuations become extremely slow below the freezing transition.

\begin{table}
\begin{center}
\begin{tabular}{cccccc}
\hline
Composition&$\theta_{CW}$&$T_g$&$\mu$&$|\theta_{CW}|/T_g$&$(|\theta_{CW}|/T_g)/\mu$\\
\hline
LuCuGaO$_4$&-69&0.4&1.73&153&100\\
LuZnFeO$_4$&-670&25&5.92&27&4.5\\
LuMgFeO$_4$&-600&16&6.02&37.5&6.2\\
LuCoGaO$_4$&-105&19&4.7&6&1.2\\
\hline
\end{tabular}
\end{center}
\caption{Observed Curie-Weiss and freezing temperatures in different RMM'O$_4$ compounds (with diamagnetic R and single magnetic transition metal), and frustration indices with normalization to effective moment size.  (Parameters from ~\cite{cava:1998} and~\cite{wiedenmann:1983}, $\mu$ is the calculated spin-only magnetic moment.) }
\label{weisstempstab}
\end{table}

\section{\label{sec:Experimental_methods}Experimental methods}

A 20 gram powder sample of \lucu was prepared by reaction of the appropriate stoichiometric quantities of Lu$_2$O$_3$ (Alfa Aesar 99.99\%), CuO (Sigma Aldrich, 99.995\%) and Ga$_2$O$_3$ (Alfa Aesar 99.999\%). CuO and Lu$_2$O$_3$ were pre-treated by heating in air for 24 hours at 900 and 1200 $^\mathrm{o}$C respectively.  The starting materials were ground together, pelletized and heated in air at 1050 $^\mathrm{o}$C for a total of 120 hours with 1 intermediate grinding.   The field cooled and zero field cooled (FC/ZFC) dc-susceptibilities were measured between 1.8 and 300 K in a field of 1000 G using a Quantum Design MPMS-7 SQUID magnetometer.  

A zero-field and longitudinal-field muon spin rotation study ($\mu$SR) was conducted between 0.06 and 50 K with fields up to 2500 G using the MuSR instrument at the ISIS pulsed muon source.    The sample was pressed tightly in a silver disk-shaped holder approximately 2 mm deep and cooled with a dilution refrigerator.

Neutron powder diffraction patterns were obtained at 70 K and 1.5 K using the D1A diffractometer at the Institut Laue Langevin (ILL) with an incident wavelength of 1.91 \AA~and standard Orange cryostat.  Polarized neutron scattering using the $xyz$ technique was performed on D7~\cite{stewart:2009} at the ILL using a wavelength of 4.8 \AA~at temperatures of 0.08, 0.5, 5 and 50 K.  Low temperatures were achieved by using a dilution refrigerator insert in a standard Orange cryostat.  The sample was contained in a copper can with $^3$He exchange gas to provide thermalization below 1 K.  The experimental background was measured using the empty sample environment and subtracted.  Corrections for detector and polarization efficiency were made by measuring standard vanadium and amorphous silica samples respectively.

An inelastic neutron scattering experiment was carried out on the chopper spectrometer IN4 at the ILL, at temperatures of 0.05, 0.5, 1.6, 5 and 50 K using an incident energy of 17 meV and similar sample environment equipment to the D7 experiment.  Detector efficiencies and elastic line position were calibrated by measuring a vanadium sample, and the scattering from the empty sample environment was subtracted.

\section{\label{sec:Results}Results}

\subsection{\label{sec:Characterization}Characterization}

We utilized powder neutron diffraction and magnetic susceptibility measurements to establish the basic quality of the sample and compare to previous studies.  The structural and magnetic characterizations of our sample agree well with previous results, particularly ~\cite{cava:1998}.

The powder neutron diffraction data show that the sample is phase pure and highly crystalline.  The Rietveld method was used to refine three structural models as described in ~\cite{cava:1998} (see also ~\cite{matsumoto:1992}).  The crystal structure has the space group $R\overline{3}m$ and  consists of double layers of edge-sharing $M^{(')}$O$_5$ triangular bipyramids connected by triangular layers of LuO$_6$ octahedra.  All three models have Cu$^{2+}$ and Ga$^{3+}$ randomly distributed on the $6c$ site, and differ with respect to the position of Lu$^{3+}$.  The Lu$^{3+}$ ions lie at the centre of distorted octahedra of O$^{2-}$ ions, directly above and below the centre of triangles of transition metal ions.  The simplest model has Lu$^{3+}$ positioned at $(0,0,0)$, midway between adjacent bilayers. The more complex models allow anisotropic motion of the Lu$^{3+}$ ion, or a shift to a lower symmetry site with random half occupation at $(0,0,\pm z)$ ($z = 0.009\pm0.0002$, corresponding to a displacement of 0.22 \AA).  In this case, the position of each Lu$^{3+}$ is thought to be controlled by the distribution of cations on the coordinating triangles of adjacent bilayers~\cite{cava:1998}.  As in ~\cite{cava:1998} the third model provides the best combination of fit and realistic parameters.  There is no indication of any structural change between 70 K and 1.5 K.  The crystal structure is illustrated in Figure~\ref{fig:LuCuGaO4xtalstructure}.

The inverse magnetic susceptibility shows a linear region at high temperature, and a downward curving form at lower temperature, in common with many frustrated magnets, even where the correlations are known to be antiferromagnetic.  We have fitted the inverse magnetic susceptibility to the Curie-Weiss law between 50 K and 300 K, obtaining $\theta_{CW} = -62.1\pm1.8$ K .  This is somewhat smaller than the previously reported value (-69 K), but in ~\cite{cava:1998} the fitting range is restricted to the temperature range 50 K to 150 K.  The effective magnetic moment, $\mu_\mathrm{eff}$, obtained from the Curie constant is $2.2 \pm 0.03 ~\mu_\mathrm{B}$ per Cu$^{2+}$ (the expected value for $S = 1/2$ is 1.7 $\mu_\mathrm{B}$ atom$^{-1}$).  The high temperature moment defined as $\mu_\mathrm{HT}\equiv\sqrt{8\chi T}=1.73 \pm 0.06 ~\mu_\mathrm{B}$ per Cu$^{2+}$ at 295 K.  The two are expected to agree in a true Curie law regime where $\theta_{CW} \ll T$, which is not really the case in these measurements.  At low temperature where the inverse susceptibility becomes non-linear, $\mu_\mathrm{HT}$ falls, indicating increasing antiferromagnetic spin correlations.  There is no discernible splitting between field cooled and zero field cooled histories at any temperature in this measurement.

One possible explanation of the enhanced low temperature susceptibility is a population of isolated or ``orphan'' spins~\cite{schiffer:1997,isoda:2008}, which could be quite reasonable in a material with strong quenched disorder.  However, a general explanation of this feature, which may be observed even in materials with very weak disorder is still debated (see ~\cite{isoda:2008} and references therein).  We have also fitted the entire temperature dependence of the inverse susceptibility with the model of orphan spins~\cite{schiffer:1997}.  This is essentially a Curie-Weiss law with an additional Curie contribution due to isolated spins, which is expected to become dominant at low temperature.  Since these spins are, in principle, isolated, they should have $\theta_{CW2} = 0$.  In this case $\theta_{CW1} = -67 \pm 2.9$ K and $\theta_{CW2} = 1.5 \pm 0.8$ K.  The ratio of effective moments implies a population of orphan spins between 19 and 16 \% depending respectively if they are assumed to individually carry moments of 1.7 $\mu_\mathrm{B}$, as for a general spin-1/2; or 2.1 $\mu_\mathrm{B}$ as the majority spins in this compound appear to do.  As mentioned above, the downward curving form of $\chi^{-1}$ is a rather general feature of many frustrated antiferromagnets  and the reality of the orphan spins~\cite{schiffer:1997} has been debated in specific cases~\cite{mendelsscgoprl,bono,Olariu:2008jk}, and in general~\cite{isoda:2008}.

\begin{figure}
	\centering
		\includegraphics[clip = true, viewport = 0.0in 4.5in 7.1in 9.15in, width=1.0\columnwidth]{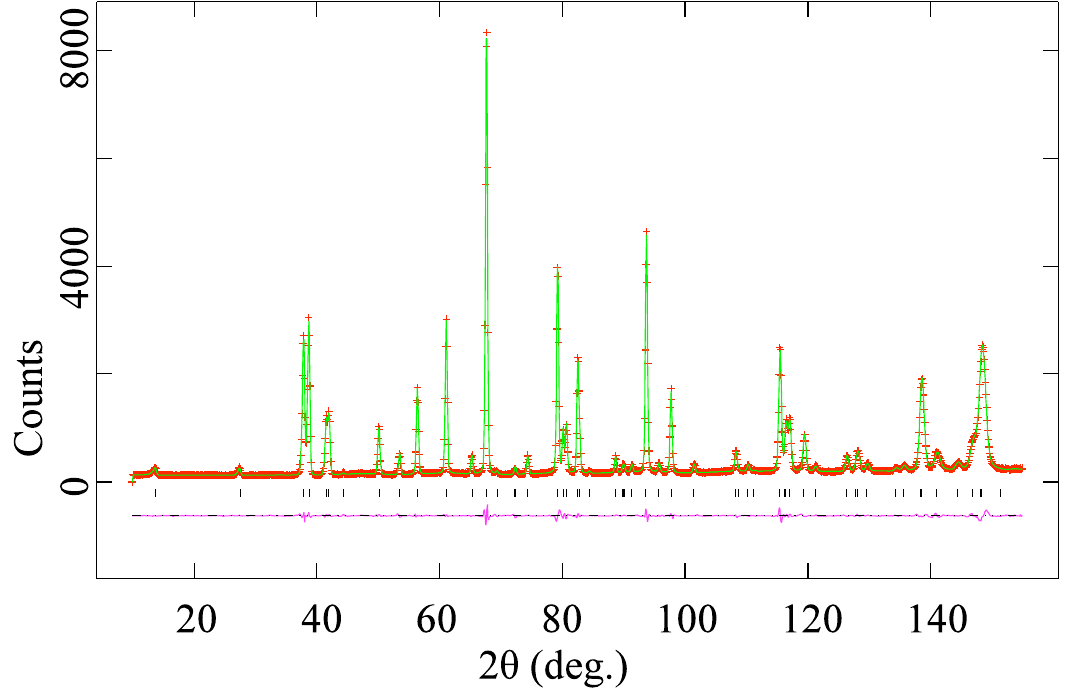}
		\caption{Neutron powder diffraction at 1.5 K.  The experimental data are the red crosses.  The fitted line (green) was obtained by using the Rietveld method to refine the model M3 of ~\cite{cava:1998}.  The corresponding structure is shown in Figure~\ref{fig:LuCuGaO4xtalstructure}.  The calculated peak positions are indicated by the black ticks and the difference between experimental and fitted profiles by the magenta line.}
\label{fig:LuCuGaO4_structure}
\end{figure}

\begin{figure}
	\centering
		\includegraphics[width=1.0\columnwidth]{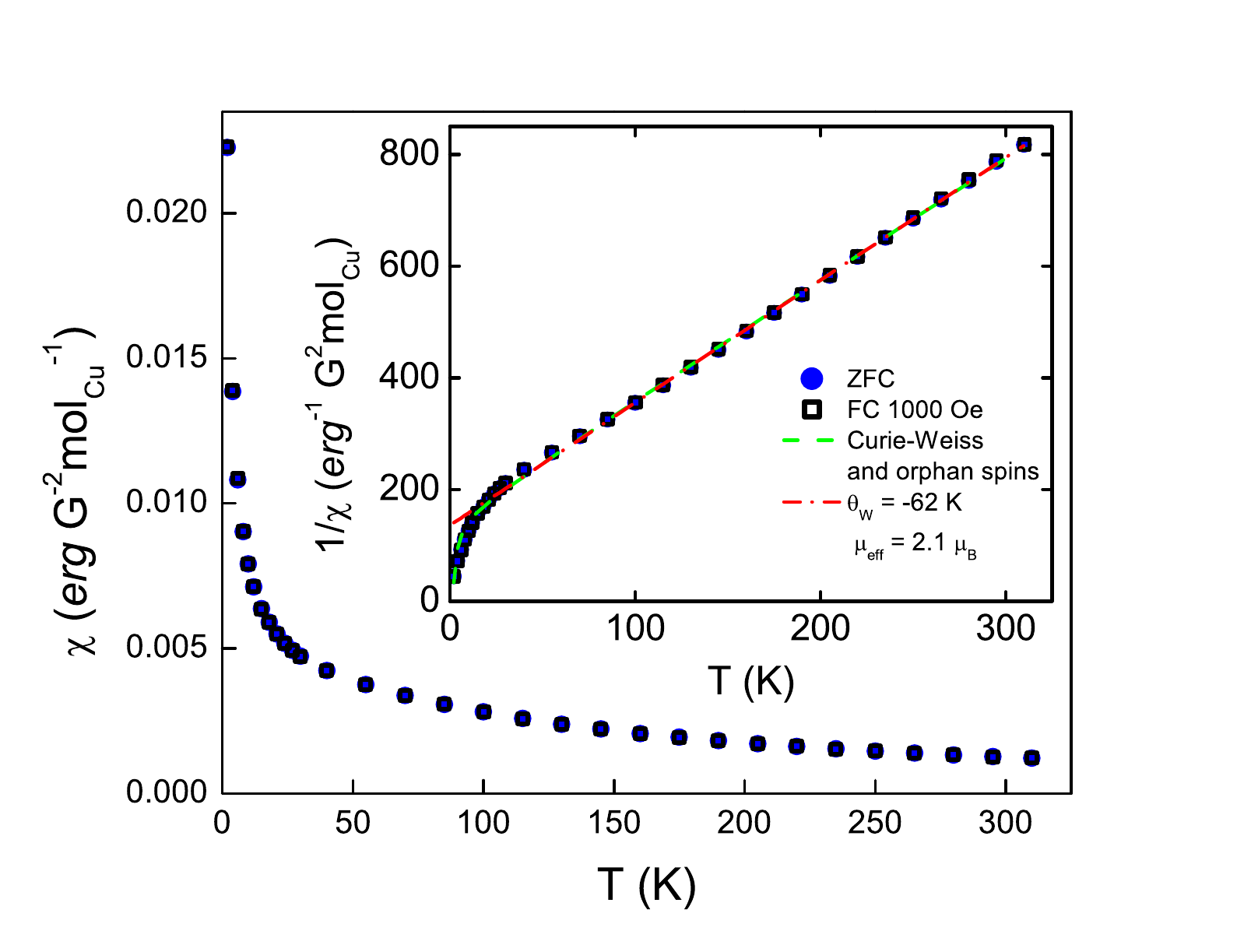}
	\caption{Susceptibility and (inset) inverse susceptibility for FC dc SQUID measurements at 1000 G. The solid line is a fit to the Curie-Wiess law between 200 and 60 K. ZFC results are identical to FC within experimental error. Also shown is the fit to a model with ``orphan'' spins~\cite{schiffer:1997}.}
	\label{fig:LuCuGaOsus}
\end{figure}

\subsection{\label{sec:muSR_res}$\mu$SR}

The development of spin correlations at low temperature was further investigated using $\mu$SR. The decay of the implanted muon polarization was measured in zero field as a function of temperature, and as a function of magnetic field at selected temperatures. $\mu$SR is sensitive to magnetic fluctuations with nuclear and electronic origins on a timescale of 10$^{-5}$ - 10$^{-7}$ seconds. In a paramagnet the electronic fluctuations are very rapid and lie outside the window of the $\mu$SR experiment.  Normally in such a situation, any observed relaxation is a consequence of the nuclear component. The nuclear component of the muon relaxation can usually be decoupled with a field of the order of $10-100$ G. We have applied longitudinal fields at several temperatures.  At 50 K, where the inverse susceptibility also shows the sample to be paramagnetic, the $\mu$SR signal is decoupled in fields of less than 100 G, suggesting the signal is entirely due to relaxation of the nuclear spins.  However, at low temperatures, where the susceptibility (and neutron scattering, see below) shows that spin correlations are developing, even a field of 2500 G is insufficient to fully decouple the relaxation.  This indicates that at low temperatures there is an increasing electronic contribution, i.e. that as the temperature is decreased, magnetic fluctuations are slowing and falling into the muon time window.

\begin{figure}
	\centering
		\includegraphics[clip = true, width=1.0\columnwidth]{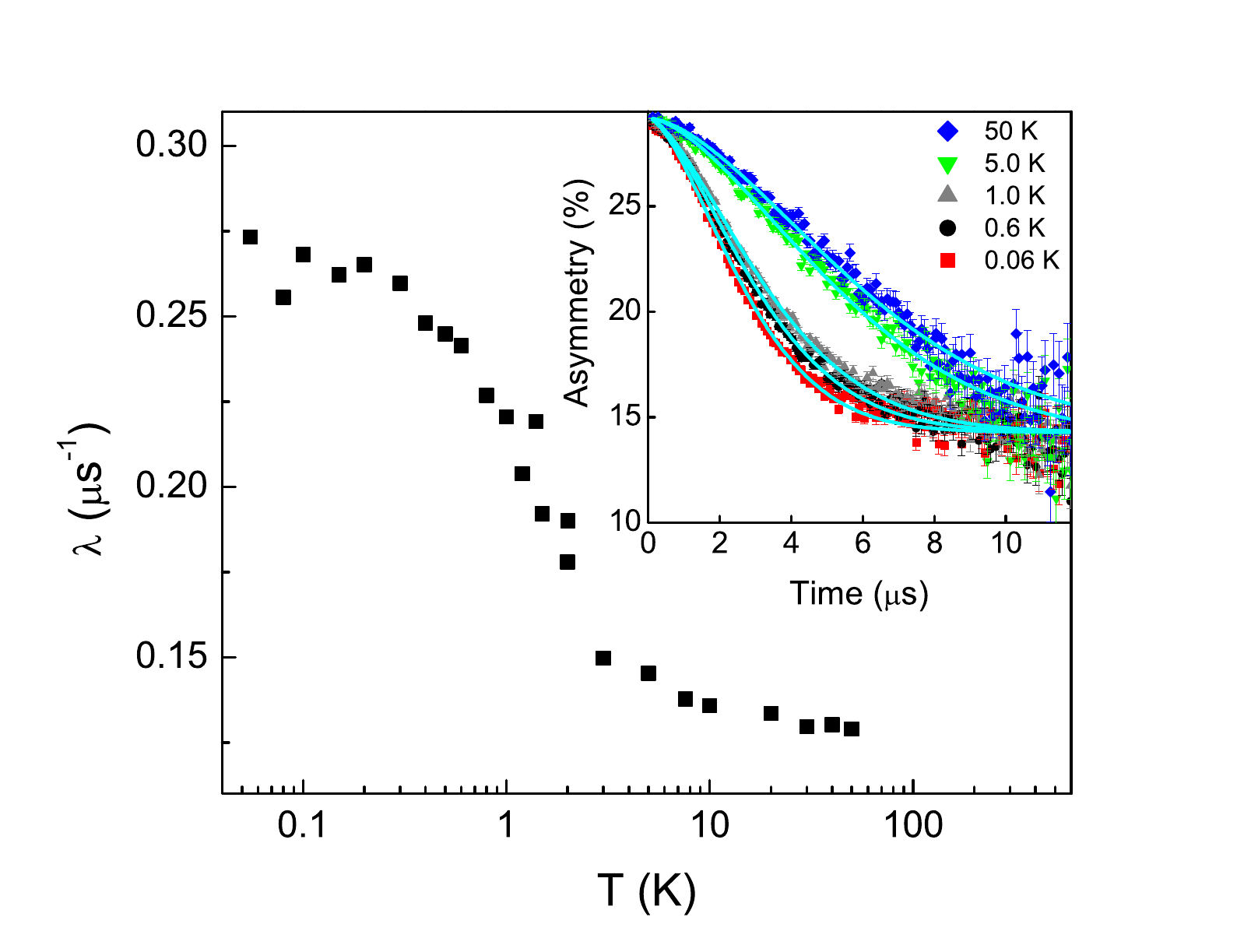}
	\caption{The muon depolarization rate in zero field as a function of temperature obtained from fits to a temperature independent nuclear response and stretched exponential electronic contribution.  Typical fitted muon responses at different temperatures are shown in the inset.  }
	\label{fig:LuCuGaO4lambdavstemp}
\end{figure}

We have fitted the zero-field muon polarization decay for different temperatures (examples are shown in the inset of Figure~\ref{fig:LuCuGaO4lambdavstemp}) with two components: a temperature independent Kubo-Toyabe function to represent the nuclear component, combined with a stretched exponential relaxation of the type $\exp ((-\lambda t)^{\beta})$, where $\lambda$ is  the muon depolarization rate.  The temperature dependence of $\lambda$ is shown in Figure~\ref{fig:LuCuGaO4lambdavstemp}.      It can clearly be seen that the depolarization rate begins to increase at $T \approx 10$ K, as the inverse susceptibility becomes non-linear.  There is a continual and relatively rapid increase until it reaches a plateau at $T\approx 0.4$ K, close to the proposed spin glass transition.  The value of $\beta$ varies by $6\%$ between $1.58\pm 0.02$ at 50 K and $1.68\pm 0.01$ at 0.055 K, and its variation is roughly proportional to $\lambda$.

\subsection{\label{sec:neutrons}Neutron Scattering}

We performed two types of experiment to investigate the development of magnetic correlations in LuCuGaO$_4$.  The $xyz$-polarization analysis technique was used to separate nuclear, spin incoherent and magnetic contributions to the total scattering~\cite{stewart:2009}, with integration of any inelastic scattering up to the incident energy (i.e. $3.55$ meV), allowing us to distinguish short range magnetic and structural correlations.  Inelastic neutron scattering using time of flight energy analysis was also used to probe the dynamical response of the system.    

\begin{figure}
		\includegraphics[width=1.0\columnwidth]{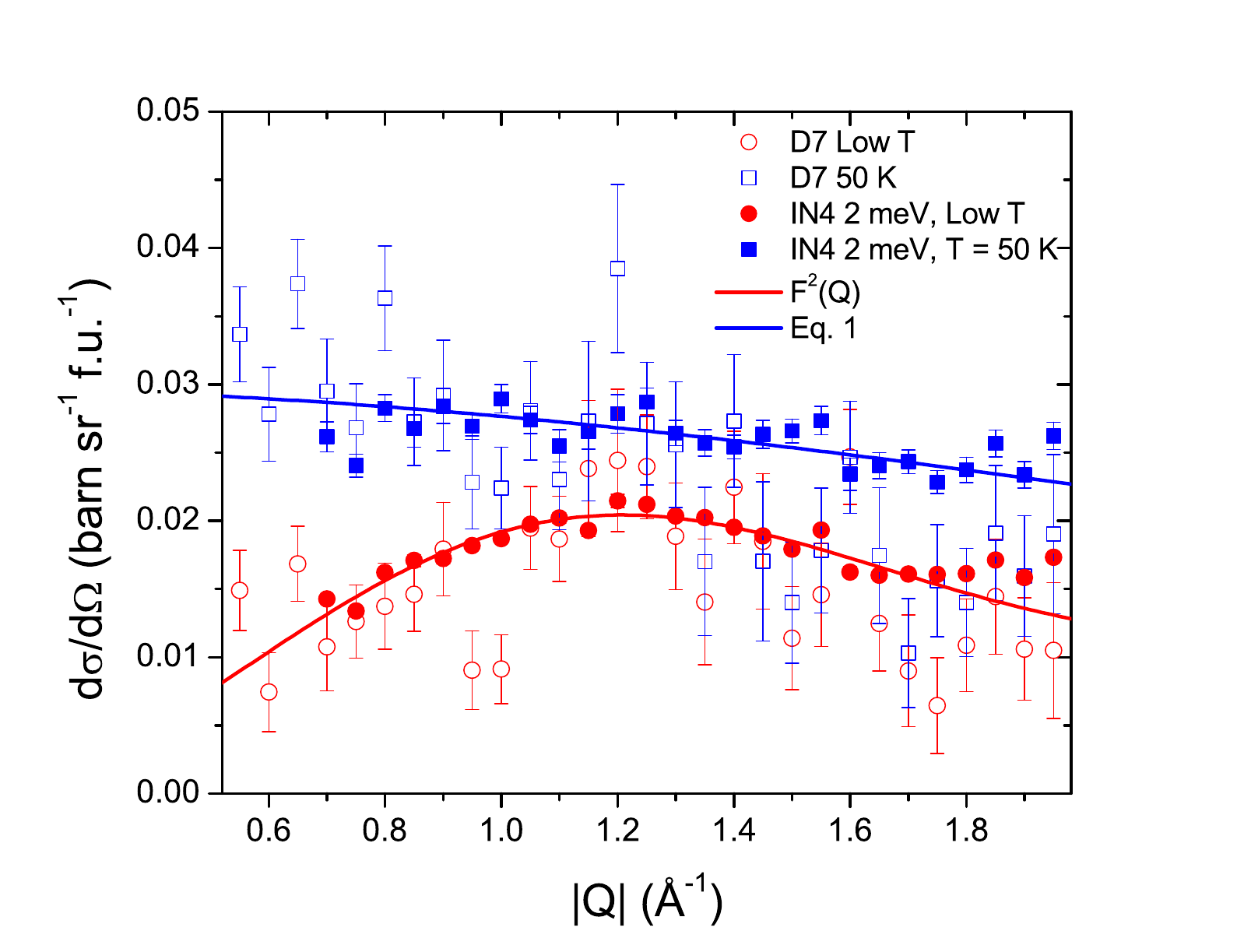}
	\caption{The development of the diffuse scattering, as measured by polarized neutron scattering (D7).  The lines show fits to the Cu$^{2+}$ magnetic form factor, and to Eqn.~\ref{AFMdimers}.  Cuts through the inelastic scattering at 2 meV are also included (IN4).  The polarization analysis experiment shows that all this scattering is purely magnetic.}
	\label{fig:LuCuGaO4D7deVries}
\end{figure}

\begin{figure}
 		\includegraphics[scale=0.8]{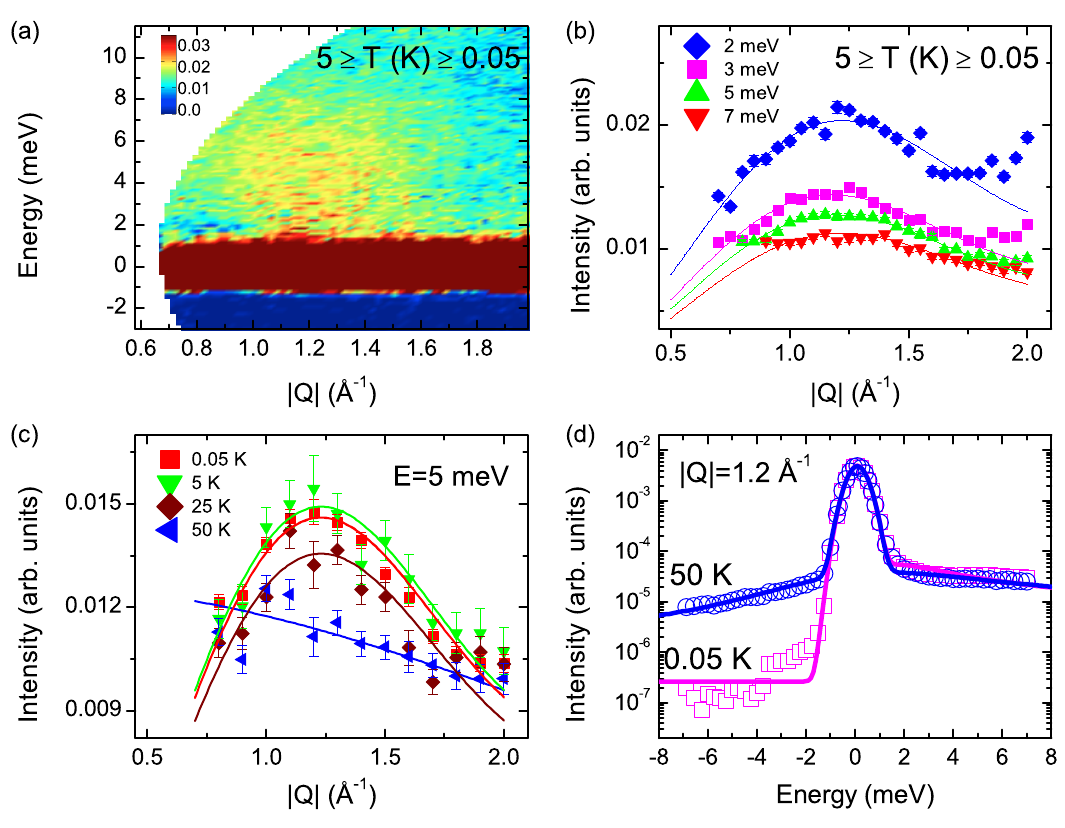}
\caption{Form and temperature dependence of the inelastic neutron scattering response.  The feature at $1.25$ \AA$^{-1}$ observed in the diffuse scattering (Figure~\ref{fig:LuCuGaO4D7deVries} can be seen to extend in energy to at least 7 meV, with no other $|Q|$-dependence (a).  It has the same form at all accessible energies (b), and at low temperatures is well fitted by Eqn.~\ref{AFMdimers}.  As the temperature is lowered, there is a redistribution of scattering intensity from a paramagnetic form-factor like response, into this feature.  At low energy, the response is still form factor like at 50 K (see 2 meV cut at 50 K in Figure~\ref{fig:LuCuGaO4D7deVries}) but the build up of this feature can already be seen at higher energies, as in (c), which shows cuts at 5 meV.  At 50 K, the energy dependence of the scattering is reasonably well described by a Lorentzian with Gaussian elastic component, but becomes non-Lorentzian at lower temperatures, as shown in (d), and also in detail in Figure~\ref{fig:Luonecut}.}				
	\label{fig:LuCuGaO4in4}
\end{figure}

Measurements using $xyz$-polarization analysis were made at temperatures selected to compare the paramagnetic regime, where the susceptibility is still described by the Curie-Weiss law (50 K), with the regime in which the inverse susceptibility becomes non-linear (5 K), and the low temperature regime around the proposed spin glass transition.  The nuclear cross section (not shown) contains only Bragg scattering, which does not change with temperature.  However, there is a clear change in magnetic scattering between 50 K and all lower temperatures studied.  At 50 K the magnetic scattering can be approximately described by the form factor of Cu$^{2+}$.  At lower temperature the form factor response at small $|Q|$ is suppressed and a broad feature appears, centered around 1.25 \AA$^{-1}$.  This can be seen in Figure~\ref{fig:LuCuGaO4D7deVries}, where the magnetic cross sections, separated from other contributions using $xyz$-polarization analysis, are shown (and also compared with constant energy cuts through the unpolarized inelastic scattering data).    There is  no significant change in the polarized neutron scattering data from 5 K down to 0.08 K, in particular between 0.5 and 0.08 K across the proposed spin glass transition (in Figs.~\ref{fig:LuCuGaO4D7deVries} and ~\ref{fig:LuCuGaO4in4} we have therefore combined the data sets from 0.08, 0.5 and 5 K).    The magnetic scattering is very weak and has therefore not been observed using conventional neutron powder diffractometers (there is no sign of this feature in Figure~\ref{fig:LuCuGaO4_structure} or the earlier study of LuCuGaO$_4$~\cite{cava:1998}).  We gain no particular understanding of the anomaly previously observed in the susceptibility and specific heat at 0.4 K~\cite{cava:1998} as there is no difference between the data measured at 0.08 K and 5 K.  This is also a feature of LuMgFeO$_4$ where no change in the neutron scattering intensity is seen across the susceptibility cusp~\cite{ikeda:1995,todate:1998:jpcm}.  

Using an expression for the structure factor of disordered, near-neighbor antiferromagnetic correlations~\cite{Stewart:2011ks},
\begin{equation}
\bigg{(}\frac{d\sigma}{d\Omega}\bigg{)} =\frac{2}{3}\frac{\gamma_n}{r_o^2} F^{2}(Q)\left(\frac{1 - Z_n\sin(Qd)}{Qd}\right)
\label{AFMdimers}
\end{equation}
with $F^2(Q)$ being the form factor for Cu$^{2+}$, $Z_n$ the number of neighbors, and $d$ the distance between the AFM correlated Cu$^{2+}$, a fit can be obtained to both the energy-integrated polarized neutron scattering data, or to cuts through the inelastic neutron scattering data at different energy transfers, as shown in Figure~\ref{fig:LuCuGaO4D7deVries}.  No appreciable difference in the fit is observed if the value of the Cu-Cu distance is  3.44~\AA~(the distance between Cu-Cu (or Cu-Ga etc) in the triangular layers), 3.04~\AA (the Cu-Cu distance across the bilayer), or a weighted average of the two (i.e. due to neighbors on both faces of the bilayer).  The position and form agrees with other members of the series where the larger moments of Fe$^{3+}$ or Co$^{2+}$ render the diffuse magnetic scattering more readily observable~\cite{wiedenmann:1983,cava:1998}.   In LuMgFeO$_4$, the dominant feature is also a peak at $Q\approx 1.25$ \AA$^{-1} $, but the diffuse scattering profile is much more detailed and correlations could be detected up to the twelfth neighbor shell.  This fitting strategy involving further neighbor correlations~\cite{wiedenmann:1983} produced unphysical values for further neighbor correlation functions in \lucuns, so we conclude that the correlations are very short ranged.  

\begin{figure}
\begin{center}
\includegraphics[trim=40 220 40 220,scale=0.4]{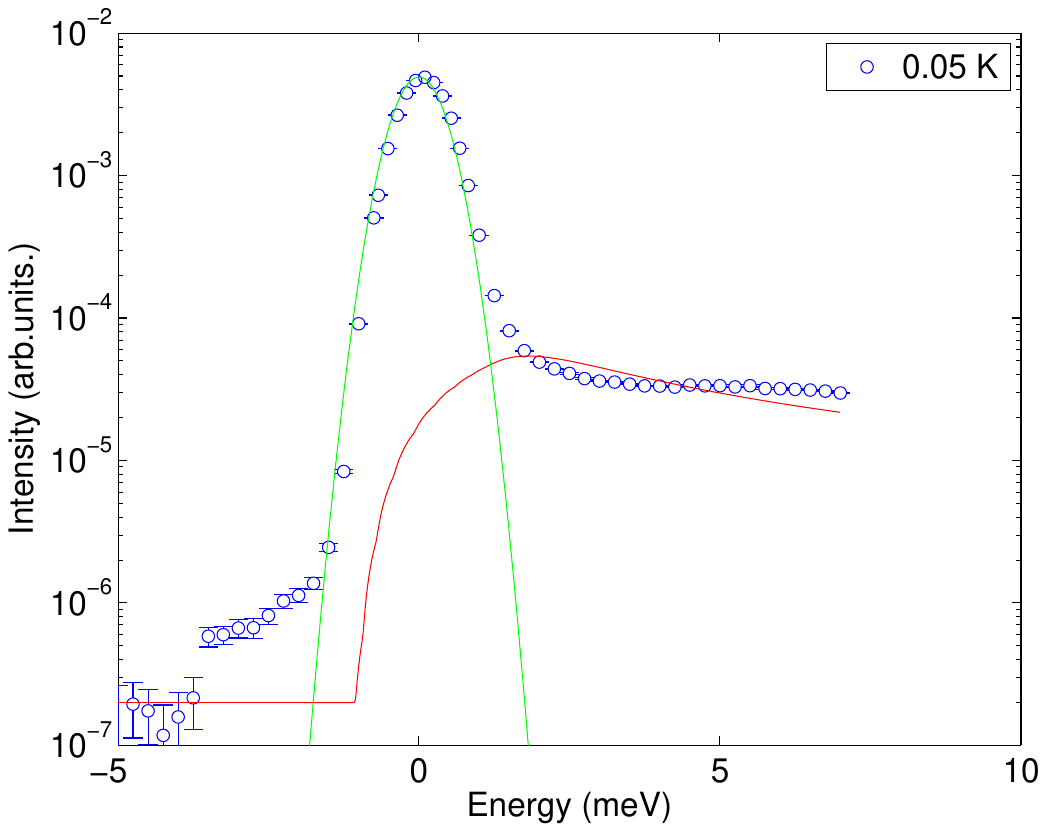}
\end{center}
\caption{Lineshape of the inelastic neutron scattering response.  Cuts through the inelastic neutron scattering data  integrated between 0.7 and 1.7 \AA$^{-1}$ were compared to a Lorentzian lineshape, which can be seen not to fit.  The Lorentzian includes detailed balance and is convolved numerically with a single Gaussian to represent the resolution.  A Gaussian incoherent contribution is also included.  The red line is the Lorentzian and the green line the Gaussian.  The shoulder may be spurious scattering from the dilution fridge insert and was not included in the fit. }
	\label{fig:Luonecut}
\end{figure}

The inelastic neutron scattering data at low temperatures show that the same feature extends upward to energy transfers of 7 meV or more.  It has proven difficult to eliminate possible phonon scattering from the copper sample container and/or sample itself at larger $|Q|$, so we confine our analysis to this feature.  Here, measurement of the empty container shows it does not contribute phonon scattering, and the polarization analysis experiment shows that the sample scattering is purely magnetic, at least up to 3.5 meV.  The feature at 1.25 \AA$^{-1}$ can be fitted by Eqn.~\ref{AFMdimers} at all energies available in this experiment, as shown in Figure~\ref{fig:LuCuGaO4in4}b.  Examination of the energy dependence of the inelastic scattering (integrated across the $Q$-range of the feature) shows that while a simple Lorentzian line shape is a reasonable description in the paramagnetic regime at 50 K, at lower temperatures, it is not (see Figure~\ref{fig:LuCuGaO4in4} and~\ref{fig:Luonecut} for examples).

The evolution of the magnetic scattering involves a redistribution of spectral weight into the feature at $1.25$ \AA$^{-1}$ as the temperature is decreased - paramagnetic intensity at low-$Q$ is suppressed, and the higher energy part of the feature becomes progressively better defined as a peak at 1.25 \AA$^{-1}$.  The redistribution of intensity takes place dominantly between 50 and 5 K, below which temperature the inelastic scattering becomes practically independent of temperature, as with the energy integrated magnetic scattering in the polarization analysis experiment.

\begin{figure}
 \begin{center}
\includegraphics[trim=70 1 60 1,scale=0.4]{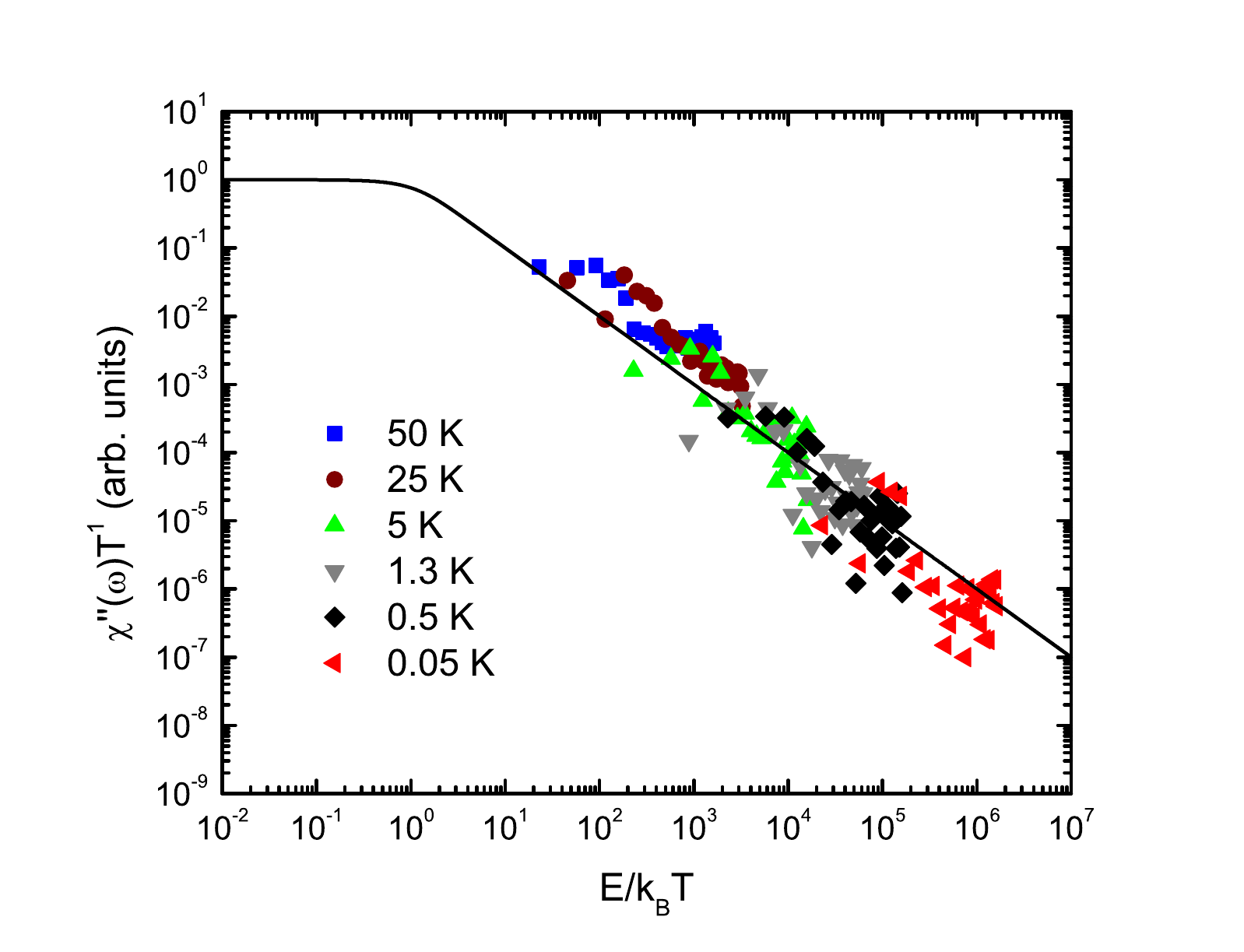}
\end{center}
\caption{Scaling collapse of the dynamical susceptibility of \lucuns.  $\chi''$ was extracted from the inelastic neutron scattering data.  When plotted in the form $\chi''(\omega)T^\alpha$, it is a universal function of $E/T$.  For the collapse to work, $0.85<\alpha<1.15$. The line is a fit to the expression $F(\omega/T)=(T/ \omega)^\alpha \tanh{(\omega / \beta T)}$ where $\beta$ is a fitting scale factor.}
	\label{fig:LuCuGaO4scaling}
\end{figure}

The dynamical susceptibility can also be extracted from inelastic neutron scattering data and examined for possible scaling collapse. In Figure~\ref{fig:LuCuGaO4scaling} we show such a comparison. Following the procedure of Helton et al.~\cite{Helton:2010ci}, the INS data was divided through by $F(Q)$, the lowest temperature subtracted as background and each data set corrected for the Bose occupation factor. The obtained $\chi''(\omega)$ was collapsed using the form  $\chi''(\omega)T^{\alpha}$ to investigate possible $E/T$ scaling.  The most useful value of $\alpha$ in the case of LuCuGaO$_4$ is $\alpha = 1$, and $\alpha$ must lie within the range $1\pm0.15$ for the scaling collapse to work.  

\section{\label{sec:Discussion}Discussion}

We first discuss the structural disorder and then the spin dynamics.  

The models fitted to the powder diffraction data treat the Cu$^{2+}$ and Ga$^{3+}$ ions as randomly distributed on the triangular bilayers, as in all previous works on related materials.  However, although the long-range ordering of two different cation species on the triangular bilayer is frustrated (see Figure~\ref{fig:LuCuGaO4topology}), in the event of a completely random distribution triangles containing entirely Cu$^{2+}$ or Ga$^{3+}$ will incur a Coulombic cost.  It is therefore possible that in these materials the cations {\it could} be correlated.  The importance of this point lies in the resulting quenched disorder configuration, which can have quite different characteristics in the case of correlated and random cations (two possibilities are briefly discussed in an appendix).

The nearest neighbor part of the Coulomb energy will be minimized by configurations of the two cations which map to the groundstates of an Ising antiferromagnet on the same lattice, as in other systems with two cation species on a frustrated lattice~\cite{anderson:1956}.  Structural diffuse scattering would be expected as a result of these short range cation correlations but since the scattering lengths of copper and gallium are extremely similar ($b_{\mathrm{Cu}} = 7.718$ barn and $b_{\mathrm{Ga}} = 7.288$ barn) direct detection of short range cation order is effectively impossible using neutron scattering.  Less direct evidence exists for the formation of short-range correlations amongst the cations.  For example, one might argue that the existence of just two defined positions for the Lu$^{3+}$ ions is due to the minimization of electrostatic interactions amongst both the intralayer and interlayer cations, but this issue unfortunately cannot be resolved in this study.  We can only conclude that the cations are certainly not ordered (long range order would presumably be manifested as lower crystal symmetry and/or distinct coordination environments).  Furthermore,  test experiments using NMR have not been able to resolve details of the Ga environment~\cite{mendelscomm}, X-ray diffraction will suffer from a lack of contrast, and the copper and gallium absorption edges that would be probed in EXAFS experiments are overlapping so further investigations of this question are not promising.

The magnetic diffuse scattering observed in the polarized neutron scattering experiments may be regarded as integrating over the quasi-elastic part of the magnetic response.  Below 5 K, the diffuse peak is temperature independent, and does not become either stronger or sharper at very low temperatures.  This implies that there is no further transfer of spectral weight into the elastic line, and therefore, that as the temperature decreases, the correlations do not become significantly more static on the timescale of this measurement.  This is also evidenced by the inelastic scattering data, which shows that by 5 K dynamical correlations on the timescale of this experiment have reached their full development, since there were no further changes in the inelastic scattering at lower temperatures.    Below 3 K however, the $\mu$SR depolarization rate increases steadily and eventually saturates.  This suggests that slower fluctuations which are observed in the $\mu$SR time window also build up.  These slower fluctuations should simultaneously appear as an increased static component in the shorter-timescale neutron scattering experiments.  That they do not may be due to the statistical quality of the neutron scattering data, or because the wavevector integration of fluctuations by neutron scattering and $\mu$SR is different.   

The three experimental probes have different timescales ranging from $10^{-5}$ s and upward for $\mu$SR, through $\sim 10^{-9}$ s (quasi-elastic neutron scattering integrated up to 3 meV), to $10^{-10}$ s for inelastic neutron scattering at $E=6$ meV.  Taken together, the data imply that fluctuations develop on many timescales.  The non-Lorentzian form of the inelastic neutron scattering at low temperature also implies that the relaxation function is not an exponential with a single characteristic time constant, and this manifests itself in the scaling collapse of Figure~\ref{fig:LuCuGaO4scaling}.   Furthermore, these fluctuations persist at low temperature, on both the $\mu$SR and inelastic neutron scattering timescales.    In ~\cite{cava:1998}, an ac susceptibility experiment which employed a single frequency (100 Hz) was reported.  $\chi''$ grew to a maximum at $T\sim 0.45$ K, below which the decrease in the susceptibility was rather weak, implying that fluctuations on this timescale also persist.  The data of ~\cite{cava:1998} and our measurements can be seen as consistent, with ever slower fluctuations appearing at lower and lower temperature.  

\lucu was previously characterized as a spin glass (on the basis of the ac susceptibility data).  However, the relaxation dynamics and scaling of the susceptibility are unlike those of a canonical spin glass.  For example, canonical spin glasses such as AgMn$_{0.5\mathrm{at.}\%}$) have been extensively investigated by $\mu$SR~\cite{keren:1996}.  There, the depolarization rate increases as the system approaches $T_g$ and the fluctuations slow down, falling into the experimental time window.  Below $T_g$ the depolarization rate then falls rapidly as the system freezes and the fluctuations become too slow for the experimental timescale.  Indeed the fluctuations typically become so slow that a spin glass in its frozen state appears static on the time scale of all neutron scattering experiments, $\mu$SR, and even susceptibility or magnetization experiments  - a pronounced peak in $\chi''$ and the well known splitting of the field cooled/zero field cooled histories result.  In contrast, in \lucu the $\mu$SR depolarization rate increases steadily as the spin correlations build up, but the system does not freeze since the depolarization rate saturates and does not fall.  A similar $\mu$SR response to that observed in \lucu is seen in a diverse variety of other materials.  In several frustrated magnets on the kagome lattice it is taken as evidence for the formation of a gapless spin liquid - examples include herbertsmithite (ZnCu$_3$(OH)$_6$)~\cite{mendels} and MgCu$_3$(OH)$_6$~\cite{Kermarrec:2011hg}, vesignieite~\cite{Colman:2011cw}, volborthite~\cite{Fukaya:2003hs,Mendels:2007gd}, and Cu(1,3-benzenedicarboxylate)~\cite{Marcipar:2009kh}.  A similar response in the FCC lattice material Ba$_2$YMoO$_6$ is attributed to the formation of a valence bond glass~\cite{deVries:2013js}, while in UCu$_4$Pd the eventual low temperature state is described as a quantum spin glass~\cite{MacLaughlin:2001ja}.

$E/T$ scaling means that the susceptibility has a general form $\chi''(\omega,T) = {\mathcal{F}}(\omega/T)$~\cite{Aronson:1995iu,fak:2008}, which contains no characteristic energy scale apart from the temperature.  Since the characteristic energy scale would give the transition temperature, and because a scaling collapse suggests the system is, in some sense, critical, the absence of a well defined transition temperature suggests a zero temperature transition.  $E/T$ scaling is therefore often taken as evidence of a nearby quantum critical point.  This scenario appears to be particularly relevant to non-Fermi liquid materials, where $E/T$ scaling is most frequently observed.  Interestingly, it has also been closely associated with disordered magnets with small spins, especially in the related experimental~\cite{HAYDEN:1991uq,KEIMER:1991ty} and theoretical~\cite{Sachdev:1992tf,Sachdev:1993tw} studies of the cuprates La$_{1-x}$Sr$_x $CuO$_4$, where a quantum phase transition between a spin glass and quantum disordered phase is thought to occur.  \cite{Sachdev:1992tf} and~\cite{Chubukov:1994ba} identify the $E/T$ scaling as  ``a general property of quantum-critical spin fluctuations, even in the presence of randomness and doping''~\cite{Chubukov:1994ba}.  Similar effects in dilute semiconductors were attributed to a temperature dependent scaling of the population of singlet pairs caused by a distribution of coupling strengths~\cite{BHATT:1982un}.

Frustrated magnets with qualitatively similar $S(\mathbf{Q},\omega)$s~\cite{fak:2008,mutka:2004,Broholm:1990tk,schweika:2007,Stewart:2011ks,devries:2009, Helton:2010ci} to \lucuns, have also been examined for possible $E/T$ scaling.  In deuteronium jarosite ((D$_3$O)Fe$_3$(SO$_4$)$_2$(OD)$_6$, kagome lattice) there is a non-Lorentzian spectrum, but no obvious $E/T$ scaling~\cite{fak:2008}; in Y$_{0.5}$Ca$_{0.5}$BaCo$_4$O$_7$ (kagome)~\cite{schweika:2007,Stewart:2011ks}  a distribution of relaxation times was observed; in herbertsmithite (ZnCu$_3$(OD)$_6$Cl$_2$, $S=1/2$ kagome) different scaling forms are observed in different frequency ranges - with $F(\omega/T)=(T/ \omega)^\alpha \tanh{(\omega / \beta T)}$ and $\alpha=0.66$ at low energies~\cite{Helton:2010ci} and independent of energy at high energies~\cite{devries:2009}; in SCGO (SrCr$_{8-x}$Ga$_{4+x}$O$_{19}$, $x=0.87$, pyrochlore bilayer), an $E/T$ scaling collapse is also possible~\cite{Mondelli:2000wi}.  Some degree of structural disorder is also associated with all these systems, particularly herbertsmithite and SCGO.   

An extensive phenomenological analysis of $\chi''(q,\omega)$ was made by Bernhoeft~\cite{Bernhoeft:2001hv}, in the context of non-Fermi liquids.  A uniform, bounded (``top-hat'') distribution of relaxation rates was considered.  In the case that the sample is characterized by a spectrum of narrow distributions of this type, which are then averaged (type 1), a non-single-exponential relaxation is expected, and it is suggested that such an averaging of relaxation rates could be associated with structural disorder.  Alternatively, there may be an intrinsic distribution of relaxation rates (type 2, i.e. a single broad top-hat distribution), which again leads to signatures of non-single-exponential relaxation.  In both cases, high frequency data collapses with $\alpha=1$, but type 1 is regarded as an uninteresting paramagnet with quenched disorder.  We would require higher resolution data to see if there is a departure from $\alpha=1$ in the low energy fluctuations.  Interestingly, for UCu$_4$Pd, both neutron scattering data with $E/T$ scaling~\cite{Aronson:1995iu}, and $\mu$SR data with qualitatively similar form to that presented here for \lucuns, exist.  The neutron scattering data was reanalyzed by Bernhoeft who proposed it to be a non-Fermi liquid of his second type (i.e. intrinsic broad distribution of relaxation times)~\cite{Bernhoeft:2001hv}, while Maclaughlin {\it et al} using $\mu$SR (to perform a different scaling collapse as a function of field, for which we do not have the relevant data) suggested it to be a quantum spin glass~\cite{MacLaughlin:2001ja}.

\section{\label{sec:Conclusion}Conclusion}

We have extended the characterization of \lucuns.  We find that at low temperatures, spin fluctuations  build up on many timescales, as evidenced by the responses falling in the different dynamic ranges of the probes employed ($\mu$SR and inelastic neutron scattering), and in qualitative agreement with previous measurements using ac susceptibility.  This leads us to a $E/T$ scaling collapse of the dynamical susceptibility of the form $\chi''(\omega)T^\alpha$, with $\alpha=1$.    These features are all inconsistent with a canonical spin glass transition to a frozen state with only very slow fluctuations. Instead these behaviours have been associated variously with quantum spin liquids, valence bond glasses, and quantum criticality, often in systems where quenched disorder is also present.  Some type of quenched disorder is certainly present in \lucuns, but we are unable to cast further light on that question in this work.  The magnetic properties of \lucu  appear more closely related to those of low dimensional magnets with quantum fluctuations than those of a canonical spin glass.  The association of the experimentally observed behaviours with multiple possible underlying physical phenomena suggests that \lucu could be an interesting model compound for the further study of quantum magnets with quenched disorder.

\ack
We are pleased to thank P. Mendels for trial NMR experiments; S Turc and O Losserand of the ILL for assistance with cryogenics on D7 and IN4 respectively; the EPSRC for funding (early parts of the work by SC and TF were supported by the grant grant EP/C534654 ``Quantum Frustration'' and carried out at the London Center for Nanotechnology); and STFC for support of beam time at ISIS and the ILL.  The research at ORNL was sponsored by the Scientific User Facilities Division, Office of Basic Energy Sciences, U.S. Department of Energy.  TF thanks M Newton of the ESRF for discussion of EXAFS, and MJP Gingras of the University of Waterloo for discussion of quantum spin glasses.

\appendix
\section{\label{sec:App}Cation Correlations}

A material such as \lucu is charge frustrated - the Ga$^{3+}$ and Cu$^{2+}$ cations are distributed on a lattice with frustrated geometry and their configurations would be expected to map to any other problem for two-state objects anti-correlated on such a lattice, i.e. antiferromagnetic spins.  The nearest-neighbour part of their Coulomb interactions will be minimized if they adopt the groundstate configurations of an Ising antiferromagnet on the same lattice.  In the case of the triangular lattice, this will mean that there will be one Ga$^{3+}$ and two Cu$^{2+}$ or vice versa per triangle.  On the bilayer, which has more interconnecting triangles at any site, and so is more frustrated, it is not clear if this condition can even be fully satisfied.  

Usually in dilute systems, one is concerned with the formation of a percolating cluster, and here it can be seen that the percolation properties could be strongly modified by the presence of correlations amongst the cations.  Since no magnetic ion would have all magnetic neighbors, dense clusters of magnetic ions could not form.    Instead a structure which consists of branching loops or chains is produced, and the anti-correlation of the cations ensures a narrower distribution of magnetic neighbor numbers.  For random occupations, the percolation threshold of an individual triangular lattice is 0.5 (for the complete bilayer it is not known), but with correlated cations, the extended nature of the cation network could ensure percolation at a lower concentration of magnetic ions.

\begin{figure}
 \begin{center}
\includegraphics[trim=200 200 200 200,angle=90,width=0.5\columnwidth]{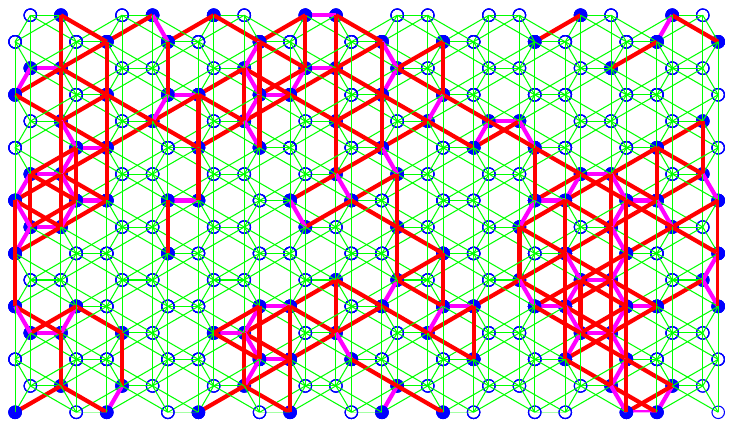}%
\includegraphics[trim=200 200 200 200,angle=90,width=0.5\columnwidth]{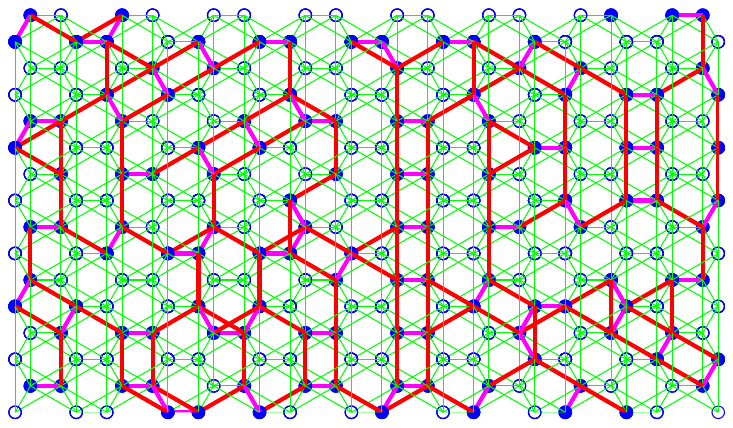}
\end{center}
\caption{Comparison of the two cation distribution scenarios discussed in the text.  Filled circles are magnetic ions (Cu$^{2+}$) and open circles are diamagnetic ions (Ga$^{3+}$).  Red lines are magnetic exchange bonds on one or other face of the bilayer, and magenta lines are magnetic exchange bonds crossing the bilayer.  Green lines complete the connectivity of the bilayer, but join either magnetic ions with their non-magnetic neighbours, or non-magnetic pairs. In the first, the cations are distributed randomly, giving the appearance of percolating clusters, in which spins can have anywhere between zero or their full complement of magnetic neighbors.  Spins can therefore be isolated, appear on branches or at cluster edges, or in the body of dense clusters.  In the second, the cations are distributed to minimize charge frustration.  They  have been mapped from a Monte Carlo simulation of antiferromagnetically coupled Ising spins on the bilayer and a few defects where triangles have zero or three neighbors can be seen.  This may be due to to the lack of a non-local update for this lattice, or possibly an intrinsic property of this lattice.  In this case, there are no isolated spins, unterminated branches, or dense clusters, since the anti-correlation of the two cations ensures that the distribution of magnetic/non-magnetic coordination numbers is narrower.}
	\label{fig:LuCuGaO4networks}
\end{figure}

In Figure~\ref{fig:LuCuGaO4networks} we illustrate the difference using two snapshots from a Monte Carlo simulation of antiferromagnetic Ising spins on the triangular bilayer (i.e. the random start and final minimized configurations).  Full characterization of the percolation properties of the clusters is beyond the scope of this work.  A non-local update for spins on this lattice is also not available, so a definitive answer to the question of whether a number of defective triangles which are fully occupied by (non-)magnetic ions are unavoidable on this lattice is also not possible.  The figures serve to illustrate  these two scenarios and show that such an effect, if present, could strongly control the magnetic properties.    

\section*{References}
\providecommand{\newblock}{}

\end{document}